\documentclass[conference]{IEEEtran}
\IEEEoverridecommandlockouts
\usepackage{cite}
\usepackage{amsmath,amssymb,amsfonts}
\usepackage{algorithmic}
\usepackage{graphicx}
\usepackage{textcomp}
\usepackage{xcolor}
\usepackage[linesnumbered,ruled,vlined]{algorithm2e}
\usepackage[]{algorithm2e}
\usepackage[]{algorithmic}
\def\BibTeX{{\rm B\kern-.05em{\sc i\kern-.025em b}\kern-.08em
    T\kern-.1667em\lower.7ex\hbox{E}\kern-.125emX}}
\begin{document}


%
\title{Understanding and Partitioning Mobile Traffic using Internet Activity Records Data - A Spatiotemporal Approach}



%

%
\author{\IEEEauthorblockN{Kashif Sultan\IEEEauthorrefmark{1}, Hazrat Ali \IEEEauthorrefmark{2},
Adeel Ahmad\IEEEauthorrefmark{1}, Kabo Poloko Nkabiti\IEEEauthorrefmark{1},  
Haris Anwaar\IEEEauthorrefmark{3}, 
Zhongshan Zhang*~\IEEEauthorrefmark{1}
}
\IEEEauthorblockA{\IEEEauthorrefmark{1}School of Computer and Communication Engineering, 
University of Science and Technology Beijing\\
30 Xueyuan Road, Haidian District, Beijing, 100083, PR China\\kashif$\_$sultan@xs.ustb.edu.cn, zhangza@ustb.edu.cn}
\IEEEauthorblockA{\IEEEauthorrefmark{2}Department of Electrical and Computer Engineering, 
COMSATS University Islamabad,\\ Abbottabad Campus, 22060, Abbottbad, Pakistan.}
\IEEEauthorblockA{\IEEEauthorrefmark{3}Department of Electrical Engineering, 
University of Engineering and Technology Lahore,\\ Kalashakako Campus, 54000, Lahore, Pakistan.}
}

%

%

\maketitle

\begin{abstract}
The internet activity records (IARs) of a mobile cellular network posses significant information which can be exploited to  identify the network's efficacy and the mobile users' behavior. In this work, we extract useful information from the IAR data and identify a healthy predictability of spatio-temporal pattern within the network traffic. The information extracted is helpful for network operators to plan effective network configuration and perform management and optimization of network's resources. We report experimentation on spatiotemporal analysis of IAR data of the \textit{Telecom Italia}. Based on this, we present mobile traffic partitioning scheme. Experimental results of the proposed model is helpful in modelling and partitioning of network traffic patterns.


\end{abstract}


%
\IEEEpeerreviewmaketitle

\section{Introduction}
The extensive development of internet of things (IoT) technologies and mobile phones suggests that next generation- networks are going in the direction where everything will be connected. An exponential increase has been seen in the use of mobile devices i.e., smart phones, tablets and smart watches. Ericsson, one of the leading telecomm company predicted the number of connected devices to be in billions in near future \cite{ericsson2011more}. The connectivity of this great number of devices will create huge data traffic. The data volume has increased by 4000 fold during last decade and this may further rise in the next generation networks (NGN)\cite{sultan2018big, sultan2017big}. Moreover, the report by CISCO tells that every month 24 Exabyte data is generated by wireless traffic which is still growing continuously \cite{Cisco}.

Useful information about the network and the behavior of subscribers can be retrieved from mobile network traces. However, network traces modeling and analysis is not trivial \cite{ak}. The information obtained by recent solutions about the modeling and analysis of these networks provide limited information and further research is needed to understand the key factors affecting the network variations. These limitations results in higher operating cost for the cellular network and dense populated areas face degradation in quality of service..

For the case of networks with ultra-density, the analysis of network traffic is very helpful for mobile operators and network subscribers \cite{dong2014inferring, shafiq2012characterizing}. Cellular network patterns, if modeled and predicted properly, can be used to further sub divide the regions accordingly i.e., the division can be done according to the traffic conditions. This optimum division of resources prove to be beneficial for subscribers in terms of quality of service and beneficial for network operators in terms of efficiently using their network resources \cite{bao2017prediction}.

Network traces have been studied extensively in the last few years to understand and model the dynamics of network. Traffic density of network has been approximated by log normal distribution by Lee \textit{et al.} \cite{lee2014spatial}. Cellular traces are analyzed spatiotemporally by Wang \textit{et al.} \cite{wang2015characterizing} and proposed that trimodal distribution is followed by cellular traces. Population density has been estimated by Deville \textit{et al.} \cite{deville2014dynamic} by analyzing call data records spatiotemporally. Moreover, traffic prediction and anomalies are proposed to be detected by analyzing network traces by Kashif \textit{et al.} \cite{sultan2018call}. Similarly, deep learning methods as reported in \cite{iqbal2018generative, bulbul20193d, ali2018speaker} can also be exploited for extraction of latent features, but are not covered in this brief review. 

Above mentioned work provided motivation to monitor and classify real mobile network traces by using artificial neural network (ANN) clustering. Internet activities record are the network traces used in this work. Spatial and temporal information is contained within the IAR data. So, we analyze the spatial and temporal data separately. Since the nature of IAR data is spatiotemporal, useful information about the network and subscribers is extracted. This information leads us to meet QoS requirement of network, and can aid in improved throughput. In the following the main contributions of this article are mentioned:

\begin{description}
  \item[$\bullet$] IAR data is analyzed of large scale cellular network. The dataset used is of Milan city, which contains internet activity for Milan. This large data allowed us in understanding and modeling urban, suburban rural areas network traffic
	\item[$\bullet$] In order to understand the spatiotemporal features of network traffic, a spatiotemporal approach is introduced. The correlation between spatial and temporal features has been investigated  to understand traffic dynamics.
	\item[$\bullet$] In order to monitor the traffic of sub regions, clustering is performed. Then, these clusters are used in ANN framework to classify the network traffic.
. 
\end{description}

The article has been structured as follows: Section \ref{sec:SystemModel} describes the proposed system model and dataset. Section \ref{sec:spatio_tempo} describes the obtained dataset spatiotemporal analysis and motivations. Section \ref{sec:clustering_predict} classification model proposed is mentioned. In the end, the conclusion is presented \ref{sec:Conclusion}.
 
\begin{figure}[b]
\centering
\graphicspath{{./figures/}}
\includegraphics[width=2.5in]{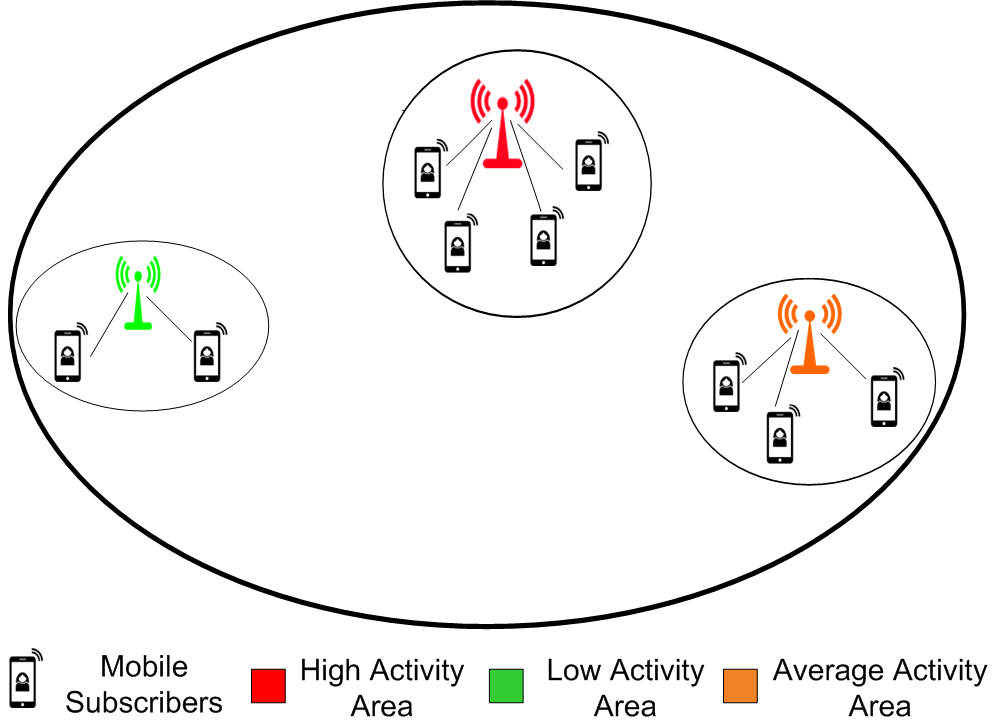}
\caption{System Architecture}
\label{fig:arc}
\end{figure}
\section{System Architecture and Dataset Description}
\label{sec:SystemModel}
The proposed system model and observed data set is described in this section. The area of Milan city is divided in to three regions namely: low activity area, medium activity area and high activity area. This categorization is done on the basis of the internet activity level and spatiotemporal characteristics. The high activity area represents the city center which include tall buildings, hospitals, shopping malls and other busy areas of city. Medium activity area comprises of city center surroundings. Whereas, low activity area comprises of area far from city center. This categorization is depicted in Fig. \ref{fig:arc}.

First of all, the internet activity records (IAR) collected is stored and processed i.e., prepared for analysis. In the next step, the spatial and temporal dependencies are observed by analyzing the spatial and temporal feature of data. Further, the IARs categorization is done in to different groups, using clustering analysis. After this step, future network traffic classification based on these clusters are done using artificial neural network (ANN). 

\subsection{Description of Dataset}
The dataset used in this article is of Milan city made public by Telecom Italia \cite{WinNT}. This dataset duration is 2 months i.e., between 1st Nov, 2013 and 1st Jan, 2014. The temporal division of the data is done in days, which is further divided in 10 minutes interval. Whereas, the spatial division is done into grids, where the entire Milan city is spatially divided in to 100$\times$100 grids space with each grid having area of 0.05 square kilometers each.

Dataset basically comprises voice calls, sms activity and internet activity. But, we only used internet activity, as these days because of rapid advancement of mobile phone technologies, internet call and messages are used more as compared to voice calls and the SMS services offered by the mobile operator. The samples of the dataset are shown in the Table \ref{tab:table3} below with cell ID, time stamp and internet activity.  
\begin{table}[h]
\renewcommand{\arraystretch}{1.2}
\caption{Telecom Italia dataset fields}
\label{tab:table3}
\centering
\begin{tabular}{|l|p{2cm}|p{3cm}|}
\hline

\textbf{Cell ID} & \textbf{Timestamp} & \textbf{Internet activity} \\
\hline
1 & 10 &57.78 \\ \hline
10 & 20 & 44.21 \\ \hline

\end{tabular}
\end{table}

\subsection{Data Preprocessing}
The dataset in raw form contains irregularities which include unrecognized numbers, misleading patterns and missing values. So, there is a need of preprocessing before analysis phase. This data preprocessing helps to improve the output quality and reduction of processing overhead. The dataset has to be aggregated in the time intervals of 10 minutes, but there are some missing timestamps. In order to overcome this issue, the activity on these missing time intervals is approximated as the average of activity in the previous and the next time interval. This estimation is represented by Eq 1. below.
\begin{figure}[b]
\centering
\graphicspath{{./figures/}}
\includegraphics[width=3in]{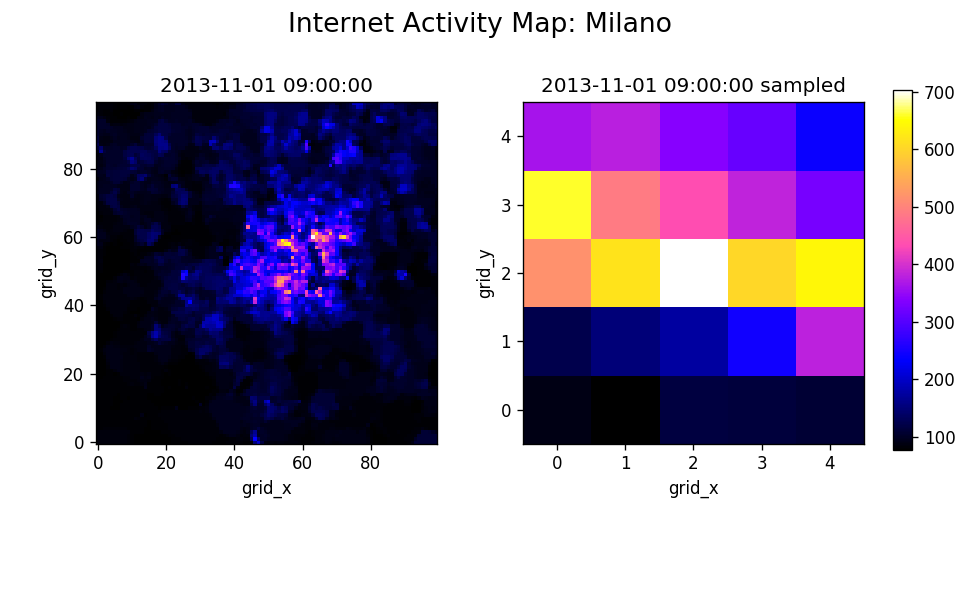}
\caption{Spatial variaton of internet activity records over Milan at a specific time instant.}
\label{fig:spatial}
\end{figure}

\begin{equation} \label{eq1}
IAR_{missing} = \frac{IAR_{t-1} + IAR_{t+1}}{2}
\end{equation}
Where ${IAR_{missing}}$ is the value at missing timestamp, ${IAR_{t-1}}$ is value of previous timestamp and ${IAR_{t+1}}$ is the value of the next timestamp.

\section{Spatial and Temporal Explorations of IARs}
\label{sec:spatio_tempo}
The preprocessed data is analyzed for spatial and temporal features separately and is mentioned in the following section.
\begin{figure*}[h]
\vspace{0cm}
\graphicspath{{./figures/}}
\includegraphics[width=\textwidth,height=6cm]{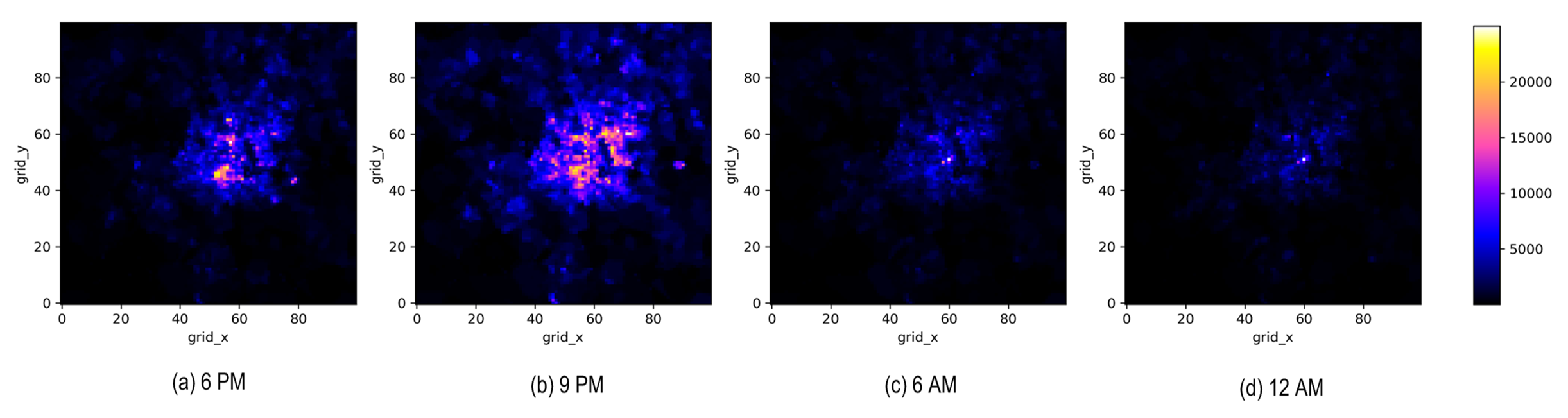}
\caption{Spatial Variation of internet activity records over the City of Milan for one day's Time intervals}
\label{fig:spatial2}
\end{figure*}
\subsection{Spatial Analysis}

While analyzing spatial feature, the temporal variations are fixed i.e., at one fixed time IARs of entire Milan city are observed. For this regard, first for one specific day, IAR data is used. After this step, the data on the three hour interval basis is aggregated. As a result, the data is further divided into 8 different time slots, over an entire period of 24 hours. 
\begin{figure}[b]
\centering
\vspace{0cm}
\graphicspath{{./figures/}}
\includegraphics[width=2in]{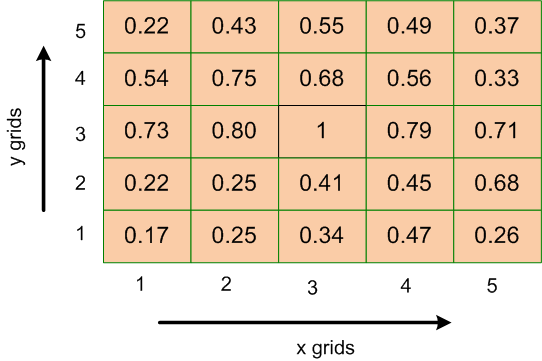}
\caption{Spatial correlation of IARs data}
\label{fig:spatial_corr}
\end{figure}

So, the activity of one whole day is observed in this way. The internet activity variation for the entire Milan city at the interval ( 6 am to 9 am) is shown in Figure \ref{fig:spatial}.In order to visualize it the 100$\times$100 grids of Milan city are downsampled to 5$\times$5. The internet activity variations can be easily observed from this 5$\times$5 grid Figure \ref{fig:spatial}. It is evident from the figure that at the center of the city internet activity is higher and it reduces when we move away from city centers. In the same way, the internet activity variation for the whole day in shown in Figure \ref{fig:spatial2}. From the Figure \ref{fig:spatial2}, the same phenomenon of internet activity variation can be observed. The internet activity spatial variation has been restricted to only four time slots because of the pages limit.
\subsubsection{Spatial Correlation}
It can be noticed that distribution of the internet activity is not uniform over entire Milan city, as different sub-regions differ in dynamics such as being a residential area or commercial area, etc. The spatial correlation variation is measured by using a widely adapted parameter, the Pearson correlation \cite{zhang2018citywide}. The Pearson correlation coefficient has been used to calculate the spatial correlation variation between the targeted grids i.e., city center grids and surrounding grids. This coefficient is expressed by the following relationship:
\begin{equation} 
	\label{eq1}
		\begin{gathered}
				\rho = \frac{cov(g_{i,j},g_{i^{'},j^{'}})}{\sigma_{i,j}\sigma_{i^{'},j^{'}}}
		\end{gathered}
\end{equation}
where $\sigma$ is the standard deviation, cov represents the covariance, $g_{i,j}$ is the target grid and $g_{i^{'},j^{'}}$ represents the grid in the vicinity of the target grid. Spatial correlation variation for the whole Milan city is shown in Figure \ref{fig:spatial_corr}. With the down sampling of 100$\times$100 grids to a scale of 5$\times$5, one cell in 5$\times$5 grid represents 400 grids.

\begin{figure}[b]
\centering
\vspace{0cm}
\graphicspath{{./figures/}}
\includegraphics[width=3in]{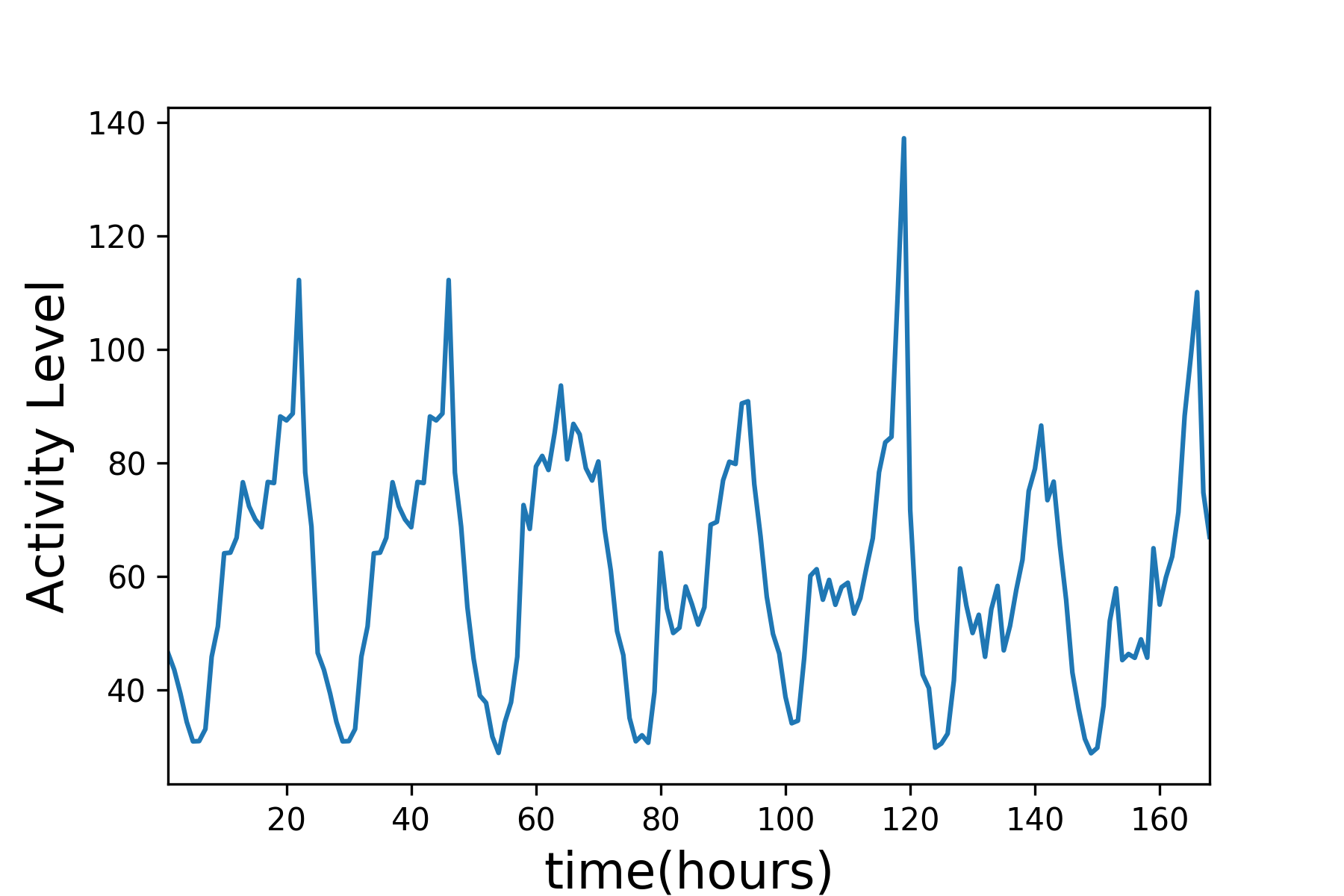}
\caption{Variation of internet activity records over a week period}
\label{fig:temporal2}
\end{figure}
\subsection{Temporal Analysis}
For this case, spatial variation is fixed and the varations in the internet activity are observed over a period  of one day, one week and one month. By fixing spatial variation, we mean observing IAR for a particular grid. From Fig. \ref{fig:temporal2} it can be seen that from midnight to morning the internet activity increases, then from noon to afternoon it decreases, and from evening to midnight it rises again. This pattern continues when observing for the case of a day, a week and a month. Fig. \ref{fig:temporal2} shows the IAR time series curve.

\subsubsection{Temporal Corelation}
\begin{figure}[h!]
\centering
\vspace{0cm}
\graphicspath{{./figures/}}
\includegraphics[width=3in]{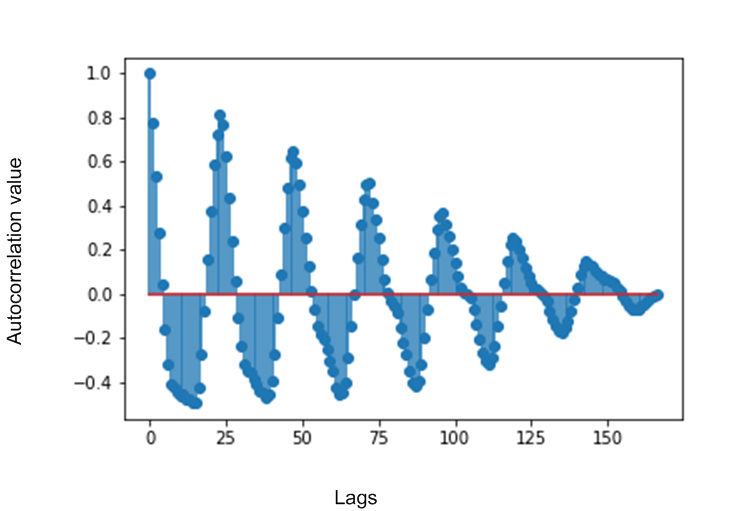}
\caption{Temporal corelation over a week period of IARs data}
\label{fig:acf}
\end{figure}

In order to check the temporal relations within IARs, the temporal correlation is measured. The temporal variation in the time series data is measured using Autocorrelation function (ACF) \cite{brownlee}. ACF measures the correlation between the current and the previous time series (also referred as lags). This type of correlation is also known as serial correlation. It is shown mathematically by the following equation 3.
\begin{equation} \label{eq1}
\gamma _\tau = \frac{\sum_{t=\tau+1}^{T} (a_{t}- \hat{a})(a_{t-\tau}- \hat{a})}{\sum_{t=\tau+1}^{T}(a_{t}- \hat{a})^2}
\end{equation}

where \textit{T} represents time steps in time series, $\hat{a}$ represents the average time series data value. The autocorrelation for 168 lags is shown in Fig. \ref{fig:acf}. The temporal dependencies of IARs are represented by these lags for the period of one week with one hour step size. It can observed from Fig. \ref{fig:acf} that hourly pattern exists in IARs as the plot of autocorrelation function shows a peak after 24 lags. 

\section{Clustering Based ANN Model}
\label{sec:clustering_predict}
The preprocessed data, after analyzed spatiotemporally is classified with respect to spatial distribution and activity level. 

\subsection{Clustering Analysis}
\subsubsection{k-means Clustering:}
Network activity level and spatiotemporal features are grouped by using clustering algorithm of machine learning’s. Clustering algorithm is used to see the variations in internet activity of entire Milan city. k-means clustering (the most popular with very low complexity) scheme is used on our dataset. The steps of k-means
clustering algorithm summarized in Algorithm 1.

\SetKw{KwBy}{by}
\begin{algorithm}[h]
\KwIn{\\ \qquad D = ${d_{1}, d_{2},........,d_{n}}$ (no. of data points)\\ \qquad k: number of clusters}
\KwOut{\\ \qquad CC = ${c_{1}, c_{2},........,c_{k}}$ (no. of cluster centroids)\\ \qquad CL = ${l_{1}, l_{2},........,l_{n}}$ (no. of cluster labels)}    
  \For{$c_{i}\gets1$ \KwTo $k$}{
    $c_{i}\gets d_{j} \in D$ 
   }
  \For{$d_{i}\gets1$ \KwTo $n$}{
    $l(d_{i})\gets argmin(d_{i},c_{j})\qquad j\in (1,...,k) $
    }
Converged = false\\

\Repeat{Converged = false}{
   \For{$c_{i}\gets1$ \KwTo $k$}
	{
	$UpdateCluster(c_{i})$
	 }
   \For{$d_{i}\gets1$ \KwTo $n$}
	{
	$minDist \gets argmin(d_{i},c_{j})\qquad j\in (1,...,k)$
  \uIf{$minDist \neq l(d_{i})$}{
    $l(d_{i}) \gets minDist$ \\
	  Converged = true 
  }

	 }
	
}

\caption{k-means algorithm}\label{alg.mainLoop}
\end{algorithm}
By the clustering analysis we obtained three types of clusters which are very low activity, medium activity and high activity clusters. The clustering approach is shown in Figure 8, with different colors representing different clusters. It can be observed from the figure that the clusters which are formed in mid of the city shows very high activities of internet. So, this is named as very high activity cluster. In the same way, the clusters which are formed near the mid of the city also shows high internet activity but lower than the mid clusters. Moreover, the clusters formed away from mid show very low internet activity. Hence, according to these internet activity clusters whole Milan city has been divided in to different groups or sub-regions.
Bayesian information criterion (BIC) and Akaike information criterion (AIC) are adapted for determining the number of components/clusters k.
\begin{figure}[h]
\centering
\vspace{0cm}
\graphicspath{{./figures/}}
\includegraphics[width=3.5in]{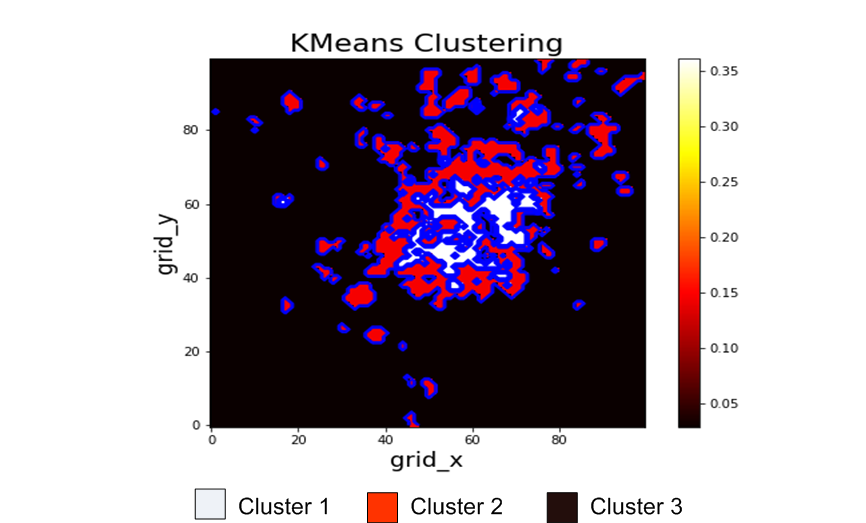}
\caption{k-means Clustering of IARs data}
\label{fig:gmm}
\end{figure}
\subsection{IARs Partitioning}
For predicting the class of IARs, clustering driven artificial neural network model is applied.  With the proposed approach, the IARs classification is converted from unsupervised to supervised multi-class classification problem. The output of the artificial neural network (ANN) model is determined with the aid of categorical distribution of IARs which is obtained from clustering analysis. ANN model classifies internet activities according to activity level and spatial characteristic.  

\subsubsection{Model Evaluation}
\begin{figure}
\centering
\vspace{0cm}
\graphicspath{{./figures/}}
\includegraphics[width=3in]{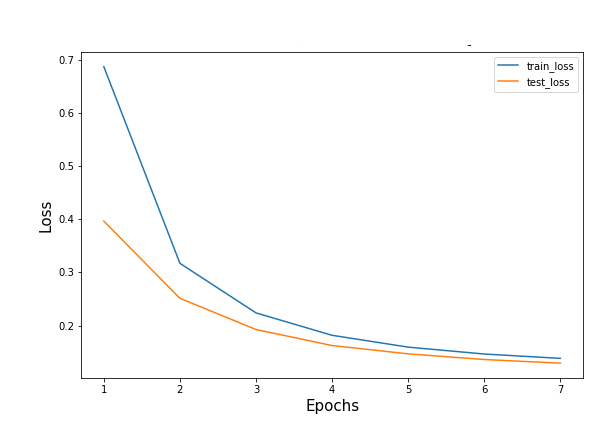}
\caption{Training and testing loss of ANN Model}
\label{fig:activity2}
\end{figure}

For observing the performance of the proposed model, loss is used as performance metrics. As in our case, the model is performing an IARs partitioning task, so categorical cross entropy is used as a loss function.
For performance visualization of the model, the training and testing losses are plotted with respect to epochs. As is observed from the simulation results shown in Fig. \ref{fig:activity2} that training losses and testing losses are decreasing as the number of epochs is increasing. It is observed from Fig. \ref{fig:activity2} that the training loss is decreasing rapidly in the first five epochs and after that reach to a stable state. The testing loss also follows the same pattern. This phenomenon proves that the model is fast converging and time efficient. 
The simulation results show that the model average loss is 18$\%$ and average accuracy is 96$\%$.

Hence, the proposed internet traffic classification model can be used for real-world applications such as efficient bandwidth allocation or multi-tier radio access network architecture. For example, the proposed IARs classification model can be used to identify the region of interest and region of non-interest. Thus, additional small cells and spectrum can then be allocated to the high activity region. In this way, QoS would be improved and energy consumption would be reduced. 
As in next-generation networks, there will be billions of connected wireless devices and a higher number of densely connected of small cells, so such type of model is heavily needed for efficient deployment of ultra-dense networks and improvement in QoS.

\section{Conclusion}
\label{sec:Conclusion}
In this work, we have presented the potential benefits of IAR data analysis in a mobile cellular network. The proposed work has provided a spatiotemporal analysis applied on mobile traffic data to extract useful pattern and help in understanding as well as monitoring the internet activity data. Results obtained for the proposed approach show that the approach is beneficial towards optimization of resource allocation task in mobile networks. Further, it provides insights into both temporal and spatial characteristics of the network IAR data. Thus, it has potential to contribute to improved performance of the network and better QoS. 
\section{Acknowledgment}
This work was supported in part by the Key Project of the National Natural Science Foundation of China under Grant 61431001, in part by
the Beijing Natural Science Foundation under Grant L172026, in part by the Key Laboratory of Cognitive Radio and Information
Processing, Ministry of Education, Guilin University of Electronic Technology. The corresponding author is Dr. Zhongshan Zhang

\bibliographystyle{IEEEtran}
\bibliography{conference_041818}

%
%

\end{document}